\documentstyle[aas2pp4]{article}

\received{30 August 1999}
\accepted{22 October 1999}

\lefthead{Grebel \& Chu}
\righthead{Hodge\,301 in 30 Dor}

\def\degr{\hbox{$^\circ$}}

\begin{document}

\title{{\em Hubble Space Telescope} Photometry of Hodge\,301:
An ``Old'' Star Cluster in 30 Doradus\footnote{Based on observations made 
with the NASA/ESA Hubble Space Telescope, obtained at the Space 
Telescope Science Institute, which is operated by the Association 
of Universities for Research in Astronomy, Inc. under NASA contract No.\ 
NAS5-26555.}
}

\author{Eva K.\ Grebel 
\altaffilmark{1,2}, You-Hua Chu\altaffilmark{3}} 

\affil{$^1$University of Washington, Department of Astronomy, Box 351580,    
Seattle, WA 98195-1580, USA}
\authoremail{grebel@astro.washington.edu}

\affil{$^2$Hubble Fellow}

\affil{$^3$University of Illinois at Urbana-Champaign, Department of 
Astronomy, 1002 West Green Street, Urbana, IL 61801, USA}
\authoremail{chu@astro.uiuc.edu}

\begin{abstract} 
We present Hubble Space Telescope Planetary Camera {\em UVI} data for the 
little-studied cluster Hodge\,301 $3'$ northwest of 30\,Doradus' central 
ionizing cluster R\,136.  The average reddening of Hodge\,301 is found to be 
$\langle E_{B-V}\rangle = 0\fm28 \pm 0\fm05$ from published infrared and 
ultraviolet photometry.  Using two different sets of evolutionary models, 
we derive an age of $\sim 20-25$ Myr for Hodge\,301, which 
makes it roughly 10 times as old as R\,136.  Hodge\,301 is the most
prominent representative of the oldest population in the 30\,Dor starburst
region; a region that has undergone multiple star formation events.
This range of ages is an important consideration for the modelling of 
starburst regions.  Hodge\,301 shows a widened upper
main sequence largely caused by Be stars.  
We present a list of Be star candidates.
The slope of the initial mass function for intermediate-mass main sequence 
stars ranging from 10 $M_{\odot}$ to 1.3 $M_{\odot}$ is found to be $\Gamma
= -1.4 \pm 0.1$ in good agreement with a Salpeter law.  There is no
indication for a truncation or change of slope of the IMF within this
mass range.  In accordance with the age of Hodge\,301 
no obvious pre-main-sequence stars are seen down to $\sim 1\,M_{\odot}$.
We estimate that up to $41\pm7$ stars with masses $>12 M_{\odot}$
may have turned into supernovae since the formation of the cluster.
Multiple supernova explosions are the most likely origin 
of the extremely violent gas motions and the diffuse X-ray emission
observed in the cluster surroundings.

\end{abstract}


\section{Introduction}

Giant H{\sc ii} regions (GHRs) are sites of active star formation and key to 
understanding starburst phenomena and active galaxies.  Their high
concentration of massive stars often make them the most prominent 
regions in distant galaxies. 
The massive stars are the primary 
providers of UV ionizing fluxes, kinetic energy, and chemically
enriched material. 
Thorough understanding and correct interpretation of GHRs require
the knowledge of their stellar content and star formation history.
This can be studied in detail only in nearby GHRs whose 
stars can be individually resolved.

With a distance of as little as $\sim 50$ kpc 30\,Doradus (30\,Dor)
in the Large Magellanic Cloud (LMC)
offers a unique opportunity to study the detailed content of massive 
stars and star formation processes in a 
prototypical giant H{\sc ii} region.  This starburst region is by far
the most luminous and massive star-forming region in the Local
Group (Kennicutt 1984) and has been called the Rosetta stone for
understanding starburst regions (Walborn 1991).  
Ground-based and space-based observations have concentrated on 
30\,Dor's central ionizing cluster, especially its core R\,136
(e.g., Parker \& Garmany 1993; Malumuth \& Heap 1994;
Hunter et al.\ 1995; Brandl et al.\ 1996; Hunter et al.\ 1997a; 
Massey \& Hunter 1998; Selman et al.\ 1999). 
While these studies yielded significant results for 30\,Dor's 
center, comprehensive understanding of the star formation history 
in 30\,Dor can only be reached by detailed studies of its other 
clusters and the surrounding field population (e.g., Lortet \& Testor 1991; 
Walborn \& Blades 1997; Rubio et al.\ 1998; Walborn et al.\ 1999).

Of particular interest for the star formation history of 30\,Dor 
is the previously neglected cluster $\sim 3'$ northwest of 
R\,136 at $\alpha_{2000} = 05^h38^m16^s, \delta_{2000} = 
-69\degr03'58''$\,:  Hodge\,301 (Hodge 1988; Lortet \& Testor 1991;
Bica et al.\ 1999).
The most luminous star in this region is the B0.5 Ia supergiant R\,132 
(Feast et al.\ 1960), which lies at a projected distance of $\sim 18''$ 
northwest 
from the cluster center ($1''\sim 0.24$ pc).  Its cluster membership is 
uncertain.
The three visually brightest stars centered on 
Hodge\,301 were found to be early A supergiants (Melnick 1985). 
Infrared surveys identified three other luminous stars in Hodge\,301
as M supergiants (Mendoza \& G\'omez 1973; Hyland et al.\ 1992). 
Another three stars in Hodge\,301 were recently classified as B-type 
giants and peculiar B main sequence stars (Walborn \& Blades 1997).
Hodge\,301 was not part of the ground-based study of 30\,Dor by Parker \&
Garmany (1993), but was observed with the Ultraviolet Imaging Telescope ({\em
UIT}, Stecher et al.\ 1992; Hill et al.\ 1993; Parker et al.\ 1998).
The massive stellar content of Hodge\,301 suggests that this cluster
belongs to a population older than R\,136
(Lortet \& Testor 1991; Grebel \& Chu 1996; Walborn \& Blades 1997).
The R\,136 cluster began to form low-mass stars 4--5 Myr ago, and its
massive stars are as young as 1--2 Myr (Massey \& Hunter 1998).

We present here our results for age, stellar content, and initial
mass function (IMF) of
Hodge\,301 based on {\em Hubble Space Telescope} ({\em HST}) observations. 
In Section \ref{sect_obs} the data and reduction procedure are described. 
In Section \ref{sect_red} we derive a mean
reddening value for Hodge\,301.  In Section \ref{sect_cont} the age of 
Hodge\,301
is determined using two different sets of evolutionary models.  The 
massive stellar
content, in particular the Be stars, are discussed.  In Section 
\ref{sect_IMF} the 
mass function of Hodge\,301 is derived and its initial massive stellar
content is estimated.  In Section \ref{Sect_context} we briefly discuss
the cluster in the framework of the star formation history of 30\,Dor, 
followed by a summary in the final section.

\section{Observations and Reduction \label{sect_obs}} 

Hodge\,301 was observed on 1995 December 17 with the Planetary Camera (PC) 
of the Wide Field Planetary Camera 2 (WFPC2; Trauger et al.\ 1994) aboard 
{\em HST} (PI: Chu). The PC chip covers an area of $36\farcs4 \times 36\farcs4$ 
with a scale of $0\farcs0455$ per pixel.  
The log of the observations is given in Table \ref{tbl_1}. 
Long exposures were split into two frames for efficient 
cosmic ray rejection. 

The aforementioned central coordinates of Hodge\,301
are from Bica et al.'s (1999) revised list of LMC
clusters.  
Our own visual estimate of the cluster center based on the Digitized 
Sky Survey (DSS)
yields $\alpha_{2000} = 05^h38^m17^s, \delta_{2000} = -69\degr04'$
(measured with the IRAF\footnote{IRAF is distributed by the National
Optical Astronomy Observatories, which are operated by the Association of
Universities for Research in Astronomy, Inc., under cooperative agreement
with the NSF.}/PROS task {\sc tvlabel}), close to the value of Bica et al.\
(1999).  We estimate that the DSS coordinates
are accurate to $\sim 2''$ in each coordinate.
The central coordinates of our PC chip
pointing are $\alpha_{2000} = 05^h38^m17^s, \delta_{2000} =
-69\degr04'03''$.
Bica et al.\ (1999) specify an apparent cluster diameter 
of $45''$ (major axis = minor axis), while we estimate an approximate
diameter of $32''\pm5''$ (from the DSS, measured where the intensity
level becomes indistinguishable from the background).
The useable field of view of the PC is $\sim 34'\times34'$. 
Thus our pointing is very close to the apparent cluster center, and the 
PC chip may be expected to enclose the majority of the cluster stars.  
Our PC field was positioned such that the brightest 
star in the region, R\,132, is outside our field of view to minimize 
overexposure effects (Figure \ref{fig_largeFoV}). 
Our field contains most of the A and M supergiants
and two peculiar B stars (Walborn \& Blades 1997).  

We used the calibrated images created by the standard {\em HST} 
pipeline processing and reduced them with IRAF
and STSDAS routines.  After cosmic ray removal, 
the images were corrected with the corresponding position- and 
intensity-dependent CTE ramps (Holtzman et al.\ 1995) and with 
geometric distortion images from the WFPC2 archives.

Stellar photometry was carried out using DAOPHOT2 (e.g., Stetson 1992) 
running under IRAF. 
Stars not found by the search routines were added by hand after visual
inspection of their profiles.  Similarly, cosmics not rejected by the IRAF
routines were removed by hand based on their profiles.  
Point-spread-function (PSF) fitting with a linearly varying Lorentzian 
PSF was found to leave the smallest residuals.
Aperture corrections were determined through aperture
photometry with Holtzman et al.'s (1995) aperture sizes and sky annuli on 
well-isolated, neighbor-subtracted stars on our short exposures.

The transformation relations given in Holtzman
et al.'s (1995) Table 7 were used to transform iteratively from the WFPC2
flight system to ground-based {\em UVI} colors.
The F656N images were not transformed.

\section{The mean extinction of Hodge\,301 \label{sect_red}}

McGregor \& Hyland (1981) list extinctions for the three red supergiants
in Hodge\,301 (Table \ref{tbl_2}).
For a mean value of $\langle A_V\rangle = 0\fm96 \pm 0\fm04$ and a ratio of 
total to selective
extinction $R = 3.5$ (Hyland et al.\ 1992) their data
yield $\langle E_{B-V}\rangle = 0\fm27 \pm 0\fm01$.  
$R = 3.5$ is valid for stars with $(J-K)>0\fm75$. 

Hill et al.\ (1993) observed 30\,Dor with the 
{\em UIT} in the B5 (1615\AA), A2 (1892\AA), A4 (2205\AA), and A5 (2558\AA)
wavebands.  
These authors used the UV and ground-based $B$-band photometry to estimate
spectral types and reddenings.  They adopted a foreground Galactic
extinction of $E_{B-V, \rm MW} = 0\fm06$ and an LMC halo extinction of 
$E_{B-V, \rm LMC halo} = 0\fm04$
from Heap et al.\ (1991).  LMC-internal extinctions were determined using 
the reddening curve of Fitzpatrick (1985: F) for 30\,Dor and the nebular 
extinction curve of Fitzpatrick \& Savage (1984: FS) together with an 
ultraviolet spectral library.  The total reddening for each of 
these stars is the
sum of the individual extinction values along the line of sight, i.e., 
 $E_{B-V, \rm tot} = E_{B-V, \rm MW} + E_{B-V, \rm LMC\,halo}
+ E_{B-V, \rm F} + E_{B-V, \rm FS}$.
Since the full width at half
maximum of the stellar images in the {\em UIT} data ranges from $2\farcs5$ to 
$3\farcs0$ not all of the {\em UIT} objects could be identified unambiguously
in our WFPC2 images. Some of the blue stars have the red supergiants 
as sufficiently close neighbors that light from both 
contributes to the measured light at {\em UIT} resolution.  
We have five stars in the PC chip in common with the {\em UIT} 
photometry (stars 247, 256, 257, 259, 264 in Hill et al.'s 
designation), which are neither close neighbors of M supergiants nor 
are known to be variable.  

Table \ref{tbl_2} contains a compilation of cross 
identifications, spectral classifications, and reddenings for stars in
Hodge\,301.  Column 1 lists the star numbers of Mendoza \& G\'omez (1973),
column 2 identifications by McGregor \& Hyland (1981), column 3 by Melnick
(1985), column 4 by Hyland et al.\ (1992), column 5 by Hill et al.\ 
(1993),
and column 6 by Walborn \& Blades (1997).  Column 7 gives spectroscopically
determined spectral types.  In column 8 spectral types derived from Hill
et al.'s (1993) {\em UIT} photometry are listed.  Note that the 
classifications based on {\em UIT photometry}
can deviate significantly from the optical
spectroscopy.  Extinctions determined
by McGregor \& Hyland (1981) are given in column 9.  The reddenings found by
Hill et al.\ (1993, see above) are listed in columns 10 (F) and 11 (FS).
The stars within our field of view are labeled in Figure \ref{fig_Befinder}.

The five above named blue supergiants and blue main-sequence stars as well
as R\,132, which is very close to the border of the PC chip, were used
to calculate the mean reddening of Hodge\,301 based on {\em UIT} 
photometry.
We find $\langle E_{B-V, \rm tot}\rangle = 0\fm29 \pm 0\fm04$ as mean 
reddening from {\em UIT} plus
Heap et al.'s (1991) extinction.   This results in a
mean extinction of $\langle A_V\rangle = 0\fm96 \pm 0\fm13$ 
for $R_{V} = 3.35$ (valid for B2\,V stars, see Grebel \& Roberts 1995).  The
uncertainties are the standard deviation of the {\em UIT} values.  The 
{\em UIT} extinction value is in good agreement with the 
results for the red supergiants from infrared photometry but shows 
larger scatter.  Note that the {\em UIT} measurements often combine
the light of several blue main-sequence stars. 
We adopt $\langle E_{B-V}\rangle = 0\fm28 \pm 0\fm05$ as mean 
reddening value for Hodge\,301.

The effect of interstellar extinction on a target star depends on the 
star's spectral energy distribution within a chosen filter.  
For a detailed description of the temperature dependence of extinction
see Grebel \& Roberts (1995), whose relations can be used to perform a
color-dependent dereddening.
The passbands of the WFPC2 broadband filters deviate from 
Johnson-Cousins $UBV(RI)_{\rm C}$ passbands.  Thus one should in principle
deredden the WFPC2 magnitudes before applying the
transformation to observed {\em UVI} colors. 
Holtzman et al.'s (1995) Table 12 gives extinction
in the primary WFPC2 filters as a function of effective temperature.  
We carried out the corresponding 
color-dependent dereddening in the WFPC2 filter system and transformed
subsequently to the observed {\em UVI} system.  A comparison of 
magnitudes resulting from color-dependent dereddening prior to 
transformation to the ground-based system and after transformation to 
this system showed only small offsets of the order of $0\fm02$. 
In the following sections we will use photometry obtained by dereddening
in the WFPC2 flight system and transforming subsequently to the 
ground-based {\em UVI} system. 

Differential reddening may be present across Hodge\,301.  Since the effects
of differential reddening are difficult to disentangle from effects of 
stellar rotation (Collins \& Smith 1985; Collins et al.\ 1991) and binarity, 
no attempt was made to deredden stars individually.  

\section{Massive Stellar Content and Age of Hodge\,301 \label{sect_cont}}

Our {\em V, (V--I)} color-magnitude diagram (CMD) shows a main sequence 
extending
over $\sim 9$ magnitudes in $V$ (Figure \ref{fig_H301_CMD}, upper panel), 
blue and red 
supergiants, and a few red giants belonging to the intermediate-age LMC
field population ($V\ga17$ mag, $(V-I)\ga0\fm5$). The {\em V, (U--V)} CMD 
(Figure \ref{fig_H301_CMD}, lower panel) 
does not extend to stars as faint as in the {\em V, (V--I)} CMD. 
Representative mean error bars are indicated in the CMDs.
Spectroscopic spectral classifications (as opposed to {\em UIT} 
classifications) of stars contained within the PC chip comprise 
the supergiants and two bright peculiar B2 stars 
(Table \ref{tbl_2}).  The main sequence shows considerable scatter at its
massive end in the {\em V, (V--I)} CMD.  As we will show in Section 
\ref{sect_Be},
this is caused mainly by Be stars, but may in part also be caused by 
differential reddening, rotation, and/or binarity.
The cluster is very well resolved in the PC images, and due to its low 
stellar density crowding should not
have much of an effect on the width of the upper main sequence.

\subsection{The Age of Hodge\,301}

We used isochrones from the Geneva group (Schaerer et al.\ 1993) and 
from the Padua group (e.g., Bertelli et al.\ 1994) at the mean metallicity 
of the young LMC population (Z=0.008) to estimate the age of Hodge\,301. 
Isochrones from the two groups have similar input physics, but 
differ in the nuclear energy generation rates assumed in particular for 
the $^{12}$C($\alpha,\gamma$)$^{16}$O reaction, 
which leads to different extensions of the blue loops (cf.\ Figure 
\ref{fig_H301_CMD}). 

Since the data were dereddened already (Section \ref{sect_red}), the remaining
ingredients for the isochrone fits are distance and age.
The distance of the Magellanic Clouds remains under debate. 
We adopt here the `canonical' value of
$(m-M)_0 \sim 18\fm5$ for the LMC (e.g., Westerlund 1997; van den Bergh 1999), 
and $18\fm55$ for 30\,Dor (Crotts et al.\ 1995; Panagia 1999).

The Geneva and Padua models were calculated for single, non-rotating stars
and should therefore reproduce the blue main sequence envelope in our CMDs.
As can be seen in Figure \ref{fig_H301_CMD} this is indeed
the case for the upper main sequence. 
Since the location of the apparent main sequence turnoff can be affected by
binaries and rotation (e.g., Grebel et al.\ 1996) and is best determined in
conjunction with spectroscopy (Massey et al.\ 1995a) we rely primarily on
the blue and red supergiants to age-date Hodge\,301.  Stars only spend
about one tenth of their main-sequence lifetime as supergiants.  Supergiants
thus allow one to constrain ages more accurately. 
Note, however, that neither set of isochrones 
can reproduce the color loci of the reddest supergiants, a shortcoming that
is also noticed in ground-based data. 

When comparing coeval Padua and Geneva isochrones with ages of the order of
$10^7$ years one finds that the main-sequence turnoff and the supergiant 
locus lie at brighter $V$ magnitudes for the Padua isochrones,
which results in higher age estimates for Padua models.  The Geneva 
isochrones indicate an age of 20 Myr for Hodge\,301, while 25 Myr is 
best compatible with the Padua isochrones.  We conservatively estimate the
uncertainty of both age determinations to be $\pm5$ Myr. 

A comparison of WFPC2 CMDs of R\,136 and Hodge\,301 is shown in Figure
\ref{fig_H301_R136}.  The R\,136 data from Hunter et al.\ (1995) 
were retrieved from the Astronomical Data Center.
The most luminous stars in R\,136 are O3, O4, and hydrogen-burning 
main-sequence WN stars with masses exceeding $120 M_{\odot}$ 
and ages of only $\sim 1 - 2$ Myr (Massey \& Hunter 1998).   Our CMDs 
indicate that Hodge\,301
is about 10 times older than the massive R\,136 cluster, which is in
good agreement with expectations from spectral classifications for its
massive stars (Table \ref{tbl_2} and Walborn \& Blades 1997).   
Hunter et al.\ (1995) and Brandl et al.\ (1996)
found pre-main-sequence (PMS) stars in R\,136 
with masses of $4 M_{\odot}$ and less (Figure \ref{fig_H301_R136}: red
stars with $M_{V,0} > 0$ mag), whose ages may range from 1--5 Myr.
In contrast, in Hodge\,301 the main sequence is well-populated down to $M_{V,0}
\sim 4$ to 5 mag.  The observed luminosity function of Hodge\,301 is shown
in Figure \ref{Fig_complete}.   
In a cluster with an age of 20--25 Myr one would expect to find 
pre-main-sequence stars at masses below $1 M_{\odot}$.
Our data of Hodge\,301 do not go deep enough to detect such stars with 
confidence.  The stars at $0\fm5 < (V-I)_0 < 1\fm2$ and $-2$ mag $< M_V <$ 1
mag are the sparsely populated red giant branch and red clump of the old field
population.

\subsection{Two Potential Blue Stragglers \label{sect_BSS}}

Two luminous blue stars ($13\fm8> V_0>13\fm6$) lie above the apparent
main-sequence turnoff but below the locus of the two A supergiants.  The 
bluer one of these two stars, WB2 in Walborn \& Blades' (1997) 
nomenclature (see Figure \ref{fig_Befinder} for its position in the cluster and 
Table \ref{tbl_3} for photometry), lies close to the end points of
the blue loops of the Geneva isochrones.  Based on its position in the
CMD one might consider it a subluminous evolved blue supergiant.
Alternatively, it may be a blue straggler candidate or a rapid rotator
seen pole-on (Collins et al.\ 1991; Pols \& Marinus 1994; 
Grebel et al.\ 1996).  This star does not show a pronounced 
H$\alpha$ excess and is not among our Be star candidates (Section
\ref{sect_Be}).  

Walborn \& Blades (1997) classify the other, redder star 
as a peculiar variable (WB9: B2 V pec var; cf.\ Table \ref{tbl_2}).  
They suggest that WB9 is a spectroscopic binary with a compact companion
whose X-ray emission is responsible for the observed He {\sc ii} 4686\AA\ 
emission.  Based on its photometric H$\alpha$ excess (see
Section \ref{sect_Be}) we identify this star as a Be star candidate.

While {\em ROSAT} High-Resolution Imager (HRI) and 
Position-Sensitive Proportional Counter (PSPC) 
data (Wang 1995; Norci \& \"Ogelman 1995) 
did not reveal X-ray point sources in the Hodge\,301 region, WB9 may be a 
Be X-ray binary, whose X-ray emission is below the detection limit of the 
existing observations.  X-ray binaries with compact companions have X-ray
luminosities ranging from $10^{34}$ to $10^{38}$ erg s$^{-1}$.  We have 
used the archival {\em ROSAT} 
HRI observation rh400779 (PI: Wang; 105 ks exposure)
to determine a 3$\sigma$ upper limit for the X-ray luminosity of WB9.
We multiplied the HRI count rate by a factor of 3 to approximate the PSPC 
count rate.  We used the visual extinction $A_V$ and the canonical gas-to-dust
ratio for the Galaxy and the LMC  (Bohlin et al.\ 
1978; Fitzpatrick 1986) to derive the H {\sc i} column density, and used it to
approximate the X-ray absorption column density.  Finally, we use the 
energy-to-count conversion factors given in {\em ROSAT} 
Mission Description (1991)
to convert from count rate to X-ray flux in the 0.1--2.4 keV band.
Both power-law models and thermal plasma emission models for temperatures
$kT$ $\sim$ 1 keV give similar conversion factors, 1 -- 3$\cdot 10^{10}$
counts~cm$^2$ erg$^{-1}$.  For a distance of 50 kpc  to the LMC, the 
3$\sigma$ upper 
limit on the X-ray luminosity of WB9 is calculated to be $\le$10$^{35}$ 
erg~s$^{-1}$.  This 3$\sigma$ upper limit is somewhat high for a 105~ks
HRI observation because WB9 is superimposed on the bright diffuse X-ray 
emission from the interior of shell 3 (Section \ref{sect_SN}, Figure 
\ref{fig_30Dor_age}).  If WB9 is indeed an X-ray binary, it must be at a 
quiescent stage or has an X-ray luminosity near the lower end of the 
range of luminosity commonly seen.

\subsection{Be Stars in Hodge\,301 \label{sect_Be}}

Be stars are non-supergiant 
(luminosity class V to III) B-type stars with H$\alpha$ emission
originating in circumstellar disks.  Free-free emission in these ionized
disks is responsible for (infra-)red and radio excess observed in Be stars.
Be stars can be distinguished photometrically from ordinary B stars through
their H$\alpha$ excess (e.g., Grebel et al.\ 1992; Grebel 1997; Keller et
al.\ 1999).  
We identified Be star candidates from images obtained in the $F656N$ filter.
$F555W$ was used as off-band filter, and $(V-I)$ as temperature indicator to
distinguish Be stars from red giants/supergiants that may also appear bright
in H$\alpha$.   

Fig.\ \ref{fig_TCD} shows the resulting two-color diagram.  The $(V-H\alpha)$
index is uncalibrated.  After adding the magnitude difference that corresponds
to the difference in filter width between $F555W$ and $F656N$ (see Grebel
et al.\ 1993), the centroid of 
blue main-sequence stars without H$\alpha$ emission is at $(V-H\alpha)\sim 0$
mag. 
We consider all stars with $(V-H\alpha) > 0\fm4$ to be Be star candidates.   

The presence of Be stars, effects of rapid rotation and/or binarity, 
leads to widened upper main sequences in young clusters
(Grebel et al.\ 1996).  The Be star fraction among main sequence B stars
appears to increase in low-metallicity environments, a possible indication
of increased rotational mixing at low metallicities (Maeder et al.\ 1999).
The reddest and more luminous Be stars (i.e., the earliest spectral types;
compare Figures \ref{fig_H301_CMD} and \ref{fig_TCD})
also tend to have the largest $H\alpha$ excess, as one might expect if
both Balmer emission and near-infrared excess originate in the 
circumstellar disks around the Be stars.
While the red excess of the brighter Be stars shows up clearly
in the {\em V, (V--I)} CMD, little red excess is seen in the {\em U, (U--V)} 
CMDs.

In order to estimate the Be/(B+Be) number ratio as a function of approximate
spectral type we used the photometric magnitude/spectral type 
calibration for B emission-line stars of Zorec and Briot (1991).  Note 
however that this gives only a crude indication of the distribution of
B and Be stars with spectral type (for a discussion see  Grebel 1997).  
Histograms
of B stars and Be stars as a function of estimated spectral type are 
presented in Figure \ref{fig_Behist}.   The Be/(B+Be) ratio peaks 
among B stars with the earliest spectral types, where about half of the
stars are Be stars.  Both total number of Be stars
and Be/(B+Be) number ratio decrease toward later spectral types and 
fainter magnitudes in accordance with other clusters of similar age 
(e.g., Grebel 1997; Maeder et al.\ 1999; Fabregat \& Torrejon 1999).
Due to decreasing Balmer emission strength toward later
spectral types and due to the variability of the Be phenomenon our
Be star counts should be considered a lower limit of the ``true''
number of Be stars.  Also, our photometric detection method is biased 
against the detection of very faint Balmer emission.

One of the Be stars (Be4 = WB6) was spectroscopically classified by Walborn
\& Blades (1997) as a B2 III giant (Table \ref{tbl_2}).  Be stars are often
rapid rotators and can appear as giants even though they may still be 
main-sequence hydrogen-burning stars (Grebel et al.\ 1996 and references
therein). 

Table \ref{tbl_3} contains coordinates and 
photometry for Be stars and other bright stars in
Hodge\,301.  Column 1 gives {\em UIT} star numbers from Hill et al.\ (1993).
The nomenclature introduced by Walborn \& Blades (1997) is listed in column 2.
Columns 3 and 4 give the coordinates of the stars.  The coordinates were
measured using the STSDAS task {\sc metric}, which gives an absolute 
accuracy of $0\farcs5$ rms in each coordinate.  The positions of the 
stars on the PC chip relative to each other have an accuracy to better than
$0\farcs005$ (Keyes 1997).  The coordinates of the stars WB1, WB8, and WB11 
were measured
on images obtained from the Digitized Sky Survey (DSS)
using the IRAF/PROS task {\sc tvlabel} (estimated accuracy $\sim 2''$ in each 
coordinate).  Column
5 denotes whether we identified a star as a Be star candidate.  For stars
outside of our field of view the entry is left blank.  Columns 6, 7, 8, and
9 give Hill et al.'s (1993) {\em UIT} photometry in the B5, A2, A4, and A5
bands (1615\AA\ to 2558 \AA, Section \ref{sect_red}).  
Their ground-based $B$-band photometry
is listed in column 11.  A comprehensive catalog of
photometry of the LMC in the {\em UIT} B1 and B5 bands was published by
Parker et al.\ (1998).  Their B1 photometry was not included in the present 
table.  Columns 10, 12, and 13 contain our 
PC photometry results.  The $K$-band photometry by McGregor \&
Hyland (1981) is given in column 14.  Figure \ref{fig_Befinder} can be used 
to identify the stars listed in Table \ref{tbl_3}. 

\section{The Mass Function of Hodge\,301 \label{sect_IMF}}

The mass function is defined as the number of stars $N$ per logarithmic mass
($M/M_{\odot}$) interval per unit area per unit time (Scalo 1986).  The 
slope of the mass function is given by $d$(log $N$) / $d$(log 
($M/M_{\odot}$)).  In this notation a Salpeter (1955) slope $\Gamma = -1.35$.   

\subsection{The Slope of the Mass Function}

Unlike the compact, rich R\,136 cluster
Hodge\,301 is a loose uncrowded cluster that is well-resolved in
our {\em HST} images; furthermore, it does not have many severely saturated 
stars.  This makes determinations of its present-day mass 
function straightforward.  Artificial star experiments indicate
that the completeness
is $\sim 100 - 93\%$ (Figure \ref{Fig_complete})
throughout the entire mass range considered in Table 
\ref{tbl_4} except for the last mass bin, where
the completeness drops 
to $\sim 75$\%.  The absence of obvious PMS stars allows us to
include most of the observed stars in the PC chip in the determination 
of the mass function.  We assume that all stars observed in the PC chip
are members of Hodge\,301 except for the few old red giants.  A comparison
with the LMC field stars found in the surrounding WF chips shows that 
the statistically expected contribution of field stars is of the order
of 7 -- 8 \% (1.5
stars for magnitudes brighter than $V_0=16$ mag, 2 stars in the 
16 mag $< V_0 <$ 17 mag
magnitude bin, 4 stars for 17 mag $< V_0 <$ 18 mag, 
8 stars for 18 mag $< V_0 <$ 19 mag, etc.).

We confine our estimation of the mass function to stars considered to be
main-sequence (core hydrogen-burning) stars based on their position in our
CMDs.  We assume that all stars lying along the main sequence are coeval
including the stars identified as Be star candidates.  Though Be stars
may spectroscopically appear as giants, we regard the ones within 
our field of view as main-sequence stars following Grebel et al.\ (1996).
While spectroscopic classifications exist for the most massive stars in
Hodge\,301, only one potential main-sequence star has a known spectral
type (WB6, see Section \ref{sect_Be}).  We will therefore assign mass bins 
based on photometry alone.  

We derive the magnitude bins corresponding to the appropriate mass intervals 
from a Geneva isochrone with an age of 20 Myr and Z=0.008.  
This procedure implicitly assumes that each star is a single, non-rotating 
star, which is certainly an oversimplification.  Logarithmic
mass intervals with the size of $\Delta$ log ($M/M_{\odot}$) = 0.1 were
chosen.  The upper limit of each logarithmic mass bin, the absolute magnitudes,
and star counts based on stars detected in both $F555W$ and $F814W$ are 
given in Table \ref{tbl_4}.  The uncertainties listed there are 
the square root of the counts.  A weighted least-squares fit results in a 
mass function slope of $\Gamma = -1.38\pm0.07$ (Figure \ref{fig_LF}, 
solid line).  The determination of the slope is based on 819 stars ranging from 
$10 M_{\odot}$ to $1.26 M_{\odot}$, covering 6.6 magnitudes in $V$. 
We note that Massey \& Hunter (1998)
found $\Gamma = -1.3$ to $-1.4$ for stars $> 15 M_{\odot}$ in R\,136 in
perfect agreement with a Salpeter slope.  

The lowest mass bin considered here (1.26 -- 1 $M_{\odot}$; 
$3\fm74<M_V<4\fm85$)
was not included in the subsequent calculation of the slope of the 
mass function due to the increasing incompleteness (see above
and Figure \ref{Fig_complete}). 
Also, toward the low-mass end of this mass bin 
PMS stars may begin to play a role.  
The highest mass bin in Figure \ref{fig_LF} 
($M_{\odot} > 10$; not included in Table \ref{tbl_4}) contains 
stars above  the apparent main-sequence turnoff.  To avoid uncertainties
introduced by supergiant mass loss and incompleteness due to evolutionary
effects, this bin was not considered in the
slope calculation, either, despite its apparently good agreement with the 
derived mass function.  

\subsection{Estimation of the Initial Massive Stellar Content and 
Mass of Hodge\,301}

Since a mass function slope derived from bona-fide main-sequence stars is
essentially an IMF slope, we can extrapolate to higher masses to estimate
the initial number of massive stars in Hodge\,301.  The resulting value
serves as a crude indication only since a sparse cluster such as Hodge\,301
would have had very few very massive stars to begin with, and we have no
way of telling up to which mass the IMF was originally populated.  Assuming
that Hodge\,301 had initially stars as massive as $\sim 120 M_{\odot}$
and a uniformly populated IMF, we find that up to $41\pm7$
stars ranging from $\approx 12 - 120 M_{\odot}$ may have turned into
supernovae since its formation.  The uncertainty
of this number was derived from
Monte-Carlo simulations to better evaluate the effects of 
small-number statistics and the stochastic nature of the upper mass cut-off
of IMFs (see, e.g., Elmegreen 1997).  

The total mass of Hodge\,301 can be calculated analytically for a chosen 
mass range and the above derived IMF slope.  Monte-Carlo simulations were
used to evaluate stochastic effects on the IMF.
Within the well-observed
bona-fide main-sequence mass bins from $1.26 M_{\odot}$ to $10 M_{\odot}$
we find a total mass of ($2388\pm71$) $M_{\odot}$.  If we assume that the 
cluster continues to follow a Salpeter IMF down to $0.4 M_{\odot}$, this adds 
additional ($2359\pm235$) $M_{\odot}$.  This value is highly speculative 
since we have not established that (1) the slope derived for intermediate-mass 
stars is valid also for low-mass stars, and (2) the Salpeter law is indeed 
valid for low-mass stars in Hodge\,301 (cf.\ Miller \& Scalo 1979).
The estimated number of low-mass stars from $0.4 M_{\odot}$ to
$1.26 M_{\odot}$ is $3647\pm362$.

Inclusion of the mass range from $10 M_{\odot}$
to $12 M_{\odot}$, which is the highest observed mass bin in Hodge\,301 and
consists in part of supergiants, adds another ($135\pm21$) $M_{\odot}$.  
This results in an
estimated present-day cluster mass of $4882\pm247$ $M_{\odot}$ from $0.4 
M_{\odot}$ to $12
M_{\odot}$.  The true uncertainty is likely larger than this.
Extrapolation from $12 M_{\odot}$ to $120 M_{\odot}$ yields
($1110\pm464$) $M_{\odot}$ and suggests
an initial stellar mass of the cluster of ($\sim 6000\pm525$) $M_{\odot}$. 

The estimate of the cluster mass from low-mass stars alone exceeds
recent estimates of the total stellar mass of Galactic open clusters
such as the Pleiades (Pinfield et al.\ 1998: $\sim 735 M_{\odot}$, $\sim 100$ 
Myr) or Praesepe (Raboud \& Mermilliod 1998: 300 -- 1330 $M_{\odot}$, 
$\sim 80$ Myr).  On the other hand, Hodge\,301 is a lot less massive than
R\,136, where some 65 stars with masses around 120 $M_{\odot}$ were
detected, while the cluster follows a normal, Salpeter-like IMF
(Massey \& Hunter 1998).

\subsection{Clues About Past Supernova Events \label{sect_SN}}

Spectroscopy of star WB9=Be1 (Walborn \& Blades 1997)
indicates that it may have a compact companion (Section \ref{sect_BSS}).
As noted by these authors, the companion may be the stellar remnant of
one of the earlier supernova events.

The interstellar medium (ISM) around Hodge\,301 gives further clues of 
a violent past.  Hodge\,301 is located within shell 3 in 30\,Dor (Cox \& 
Deharveng 1983).  Shell 3 has a diameter of 120 pc, which is comparable 
to those of typical superbubbles formed by OB associations or clusters 
in the LMC.  Shell 3 nevertheless has two remarkable 
features: bright diffuse X-ray emission (Wang \& Helfand 1991; Wang 1999,
see his Figure 3 particularly) and highly supersonic gas motions 
(Meaburn 1988).  Observations of 30\,Dor's kinematics show that 
shell 3 hosts multiple expanding substructures and that some of the 
substructures have the most extreme expansion velocities, approaching 
300 km~s$^{-1}$, in 30\,Dor (Chu \& Kennicutt 1994, see their Figures 7c 
and 7d).  Both the diffuse X-ray emission and violent motions of shell 3 
are characteristic of supernova remnants, and the multiple centers of
expansion further indicate multiple supernova explosions.  These physical
properties thus fully support a dynamic interaction between Hodge\,301 
and the ambient ISM in 30\,Dor.  To study the details of this interaction,
we have mapped the velocity structure of the WFPC2 field using long-slit
echelle spectra; these data will be reported in another paper (Chu et al.,
in preparation).
 
\section{Hodge\,301 in the Context of 30\,Doradus \label{Sect_context}}

Walborn \& Blades (1997) identify four different spatial and temporal
structures in the starburst region of 30\,Dor.  The Hodge\,301 cluster
is the most prominent representative of the oldest component.  
Other recognizable members of this oldest population include a few massive 
M and A supergiants found mainly to the east and south of the central 
ionizing cluster (e.g., Westerlund 1961;
Mendoza \& G\'omez 1973; Hyland et al.\ 1978; McGregor \& Hyland 1981; Parker
1993).  In Figure \ref{fig_30Dor_age} we have marked stars with spectroscopic
classifications that trace the different subpopulations in 30\,Dor.  The
distribution of K, M, and A stars is representative of the oldest 
population are shown in the upper left panel of this figure. 

Recent {\em HST} NICMOS observations of 30\,Dor (Walborn et al.\
1999) confirmed embedded protostars (Walborn \& Blades 1987;
Hyland et al.\ 1992; Rubio et al.\ 1992),  and led to the detection of
previously unknown small embedded clusters and Bok
globules with forming stars.  Together with the younger components listed
by Walborn \& Blades (1997), this demonstrates that star formation in
the nearest extragalactic starburst region has occurred repeatedly
over the past 20 -- 25 Myr and is still continuing.

Hyland et al.\ (1992) suggested that the massive stars observed today as 
red supergiants in the vicinity of the R\,136 cluster may have triggered 
the star formation in its birth cloud making 
it commence from many directions.  The multiple H{\sc ii} shells in the 30\,Dor
periphery, which show pronounced diffuse X-ray emission (Wang \& Helfand
1991; Wang 1999; and upper left panel in our Figure \ref{fig_30Dor_age}) 
and which are likely the result of repeated 
supernova explosions seem to give further support to an ``outside-in''
scenario of star formation.  The most recent star formation is occurring 
in the dense filaments surrounding R\,136 (examples are marked in the lower
right panel of  Figure \ref{fig_30Dor_age}).  However, there is no clear
evidence of sequentially triggered star formation, and it is conceivable
that the repeated star formation events occurred {\em sua sponte} in a
clumpy, increasingly turbulent giant molecular cloud.

\section{Summary and Concluding Remarks \label{sect_conc}}

We have analyzed {\em HST} WFPC2 data of the sparse cluster Hodge\,301 in
30\,Dor, which lies $3'$ northwest of the massive central ionizing
cluster R\,136.  The published spectral classifications of the supergiants
in Hodge\,301 indicate already that this cluster is not as young as
R\,136.  Depending on the evolutionary models used our photometry yields
an age of $25\pm5$ Myr (Padua isochrones) or $20\pm5$ Myr (Geneva isochrones),
which is $\approx$ 10 times as old as the age of R\,136 as derived from its
massive stars. 

Hodge\,301 contains
Be star candidates, which we identified based on their H$\alpha$ excess.
The highest Be star fraction is found among the earliest B spectral 
types at the main-sequence turnoff region, as is typical for clusters
in this age range (Grebel 1997).  We provide a finding chart and a table 
with equatorial coordinates and photometry of the Be star candidates.
Two stars above the apparent main-sequence
turnoff but below the locus of the A supergiants may be blue stragglers. 
One of them was identified as a possible binary with a compact companion
(Walborn \& Blades 1997) and shows H$\alpha$ emission in our data.  This
star may be a Be X-ray binary.  We derive a 3$\sigma$ upper
limit on the X-ray luminosity of WB9 of $\le$10$^{35}$
erg~s$^{-1}$ in the {\em ROSAT}  0.1--2.4 keV band. 

We derive a mass function for the main-sequence stars in Hodge\,301
spanning a mass range from 10 to 1.26 $M_{\odot}$ and find a 
slope of $\Gamma = -1.38 \pm 0.07$.  Thus 
Hodge\,301 has a normal mass function slope in excellent agreement with a 
Salpeter slope and with 
the results of Massey et al.\ (1995a) and Hunter et al.\ (1997b) 
for young associations and clusters in
the Magellanic Clouds ($\Gamma = -1.3\pm0.3$).
The slope is slightly steeper than what Hunter et al.\
(1995; 1996) and Selman et al.\ (1999) found for the intermediate-mass 
stars in R\,136 (slopes ranging from $\Gamma = 
-1.0 \pm 0.1$ to $-1.25 \pm 0.05$), 
and steeper than the mean slope in the Milky Way clusters and 
associations  ($\Gamma = -1.1\pm0.1$; Massey et al.\ 1995b). 
As discussed by Hunter et al.\ (1997b) and Sagar \& Griffith (1998),
deviations of this magnitude may
be more indicative of the overall uncertainties in the derivation of 
a mass function and of the corrections applied 
than of a significant change of slope.
Sirianni et al.\ (1999) suggest that the IMF of R\,136 shows
a clear flattening below $3 M_{\odot}$.  In Hodge\,301 no such flattening
is observed down to $\sim 1 M_{\odot}$. 
We estimate the total present-day mass of Hodge\,301 to be 
$4882\pm247$ $M_{\odot}$ from $0.4 M_{\odot}$ to
$12 M_{\odot}$ if a Salpeter function with a single slope is indeed
valid over this entire mass range.

Extrapolating Hodge\,301's mass function to $120 M_\odot$ we find that
as many as $41 \pm 7$ massive stars may have evolved into supernovae
since the cluster was formed.  This is consistent with the 
highly supersonic gas motions and the diffuse X-ray emission in
Hodge\,301's surroundings, which indicate multiple
supernova explosions.

30\,Dor is considered a prototypical starburst region.
Its complexity illustrates the importance 
to consider multiple generations of star and cluster formation
when interpreting distant,
unresolved starburst regions.  Their integrated light may be the result of
repeated, distinct star formation
episodes that may have spanned a few $10^7$ years rather than just one
instantaneous star formation event.  Their gas kinematics and velocity
dispersion depends on their star formation history and the energy feedback
from massive stars.  Apart from turbulence, thermal broadening and  
a gravitational component contribute (e.g., Yang et al.\ 1996).  Depending 
on the evolutionary stage and
star formation history of a giant HII region, the proportionality
of these three components can be different.
The complex evolutionary history and spatial structure
should be taken into account in the
modelling of GHRs and starbursts, as Thornley et al.\ (1999) have attempted 
for 30\,Dor.

\acknowledgements
We thank Wolfgang Brandner for many useful discussions and our referee,
Joel Parker, for helpful comments.
Nolan Walborn kindly made available spectral classifications prior to
publication.  We also wish to acknowledge the use of tabulated data
distributed by the Astronomical Data Center at NASA's Goddard Space 
Flight Center and of {\em ROSAT} HRI data obtained from the High Energy 
Astrophysics Science
Archive Research Center (HEASARC), provided by NASA's Goddard Space Flight 
Center.  Some of the coordinates
were measured on an image obtained from the Digitized Sky Survey.
The Digitized Sky
Surveys were produced at the Space Telescope Science Institute under U.S.\
Government grant NAG W-2166. The images are based on
photographic data obtained using the UK Schmidt Telescope.
The UK Schmidt Telescope was operated by the Royal Observatory Edinburgh,
with funding from the UK Science and Engineering Research Council (later
the UK Particle Physics and Astronomy Research Council), until 1988 June,
and thereafter by the Anglo-Australian Observatory. 
We also acknowledge extensive use of NASA's Astrophysics Data System 
Abstract Service.  

This research was supported by NASA through the 
grant STI6122.01-94A.  EKG acknowledges support by the German Space Agency 
(DARA) under grant 05 OR 9103 0 and by NASA through grant 
HF-01108.01-98A from the
Space Telescope Science Institute, which is operated by the Association of
Universities for Research in Astronomy, Inc., under NASA contract NAS5-26555.

\clearpage

\figcaption[grebel.fig1.ps]{Location of the WFPC2 field in 30 Doradus.  
Hodge\,301 is
contained in the field of view of the PC chip.  The long sides of the
WFPC2 footprint are $\sim 80''$ long.  The underlying image is a 
composite of a broadband {\em R} and an [S{\sc ii}] image obtained
with the NTT and EMMI in 1995.  The location of the B0.5 Ia star R\,132,
which lies just outside of the PC chip, is indicated as well.
\label{fig_largeFoV}}

\figcaption[grebel.fig2.eps]{Left: A composite image 
({\em F336W, F555W, F814W}) of Hodge\,301 
(PC chip).  Right:  The same image with star identifications.
Be star candidates identified
through their H$\alpha$ excess are marked by circles.  Star numbers from 
Walborn \& Blades (1997) are prefixed by WB and are marked by diamonds unless
they were identified as Be stars.  The stars from the {\em UIT} study of Hill 
et al.\ (1993) that do not have other identifiers are marked by squares.  
See text and Tables \protect{\ref{tbl_2}} and \protect{\ref{tbl_3}} for more
information.  The three concentric circles indicate apertures of $1''$, 
$2.5''$, and $3''$ to illustrate the crowding problems in the {\em UIT} data
(Section \protect{\ref{sect_red}}).
\label{fig_Befinder}}

\figcaption[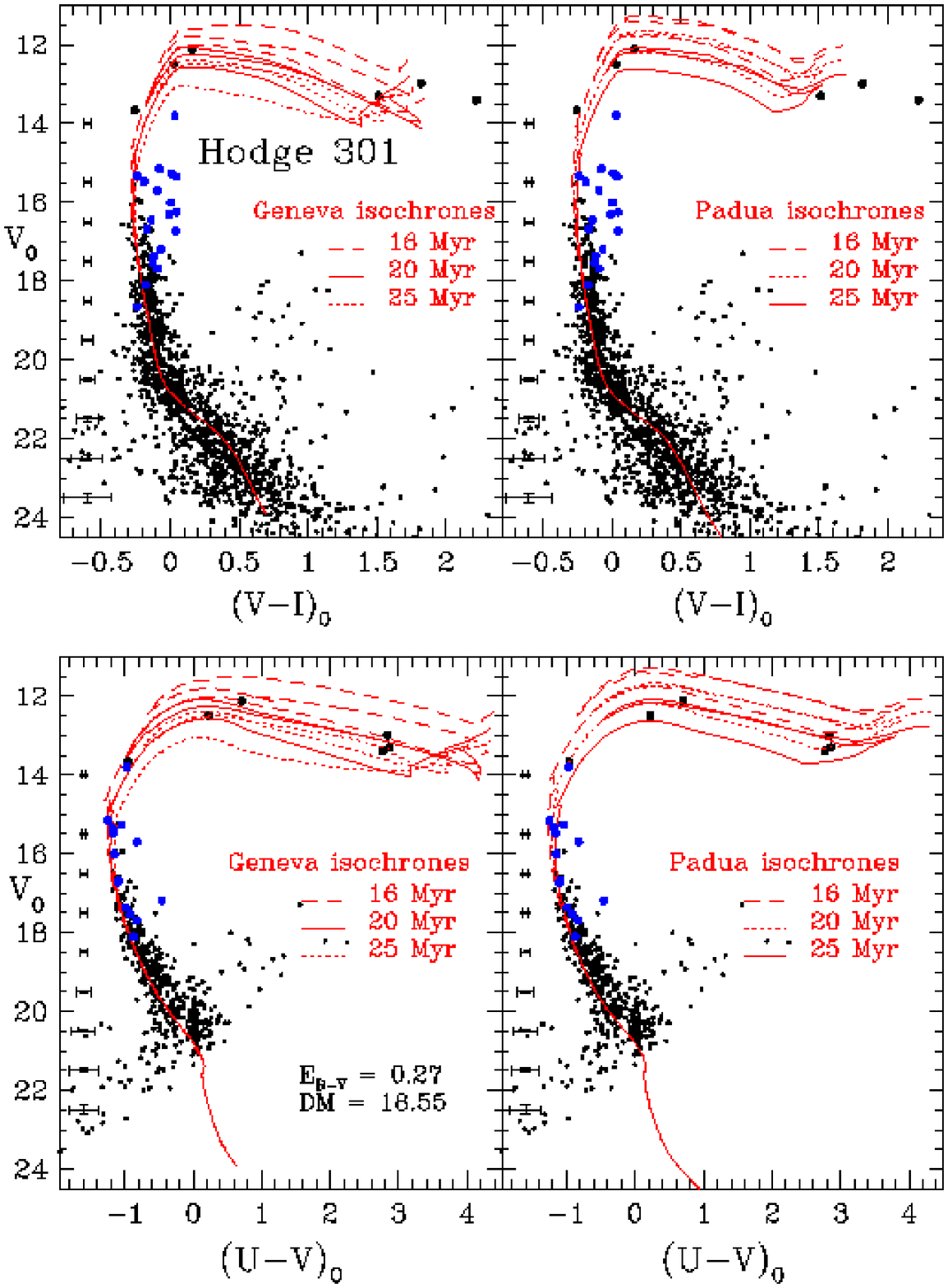]{$(V-I)_0$, $V$ (upper panel) and 
$(U-V)_0$, $V$ (lower panel)
color-magnitude diagrams (CMDs) of Hodge\,301 with Geneva (left) and Padua
isochrones (right) at Z=0.008.  B emission-line stars are indicated through
grey dots (Section \protect{\ref{sect_Be}}).  Stars brighter than 
$V=14$ mag were plotted as fat dots to retain
legibility.  Representative error bars are plotted in the left-hand side
of each CMD.  An age of 20--25 Myr (solid lines) seems best compatible with
the isochrones.
\label{fig_H301_CMD}}

\figcaption[grebel.fig4.eps]{Comparison between the color-magnitude 
diagrams of R\,136
(data taken from Hunter et al.\ 1995 and dereddened using their values) 
and Hodge\,301.  Both datasets were obtained with WFPC2 and have the same 
exposure times.  Geneva group isochrones for Z=0.008
are overplotted as solid lines.  
A pre-main-sequence (PMS) track is shown for R\,136 as a dashed line.
PMS stars can be seen in the R\,136 diagram at $M_{V,0} \sim 0\fm3$ and 
fainter magnitudes.  At $M_{V,0} \ga 3\fm5$  the contribution from PMS stars
becomes dominant.  No such trend is seen in the much older Hodge\,301.  
A sparsely populated red giant branch and red clump of the old field 
population can be seen in the Hodge\,301 CMD at $0\fm5 < (V-I)_0 < 1\fm2$
and $-2$ mag $ < M_V <$ 1 mag.
\label{fig_H301_R136}}

\figcaption[grebel.fig5.eps]{Magnitude histogram for the PC data of
Hodge\,301 (solid line).  The
incompleteness derived from artificial star experiments is indicated by
long dashed lines in the faintest magnitude bins.  No attempt was made to 
correct the last bin, which is severely affected by incompleteness.
The luminosity function of Hodge\,301 is compared
to a luminosity function (LF) derived by Hunter et al.\ (1995; adapted from
the completeness-corrected upper solid line in their Figure 19) for R\,136.
The R\,136 LF ends where Hunter et al.\ (1995) considered the incompleteness
too large for reasonable corrections. 
The LF of Hodge\,301 extends to fainter magnitudes due to significantly less 
crowding in this sparse cluster as compared to the R\,136 0.7 -- 1.4 kpc
annulus.  At $M_{V,0} \sim 0\fm3$ pre-main-sequence stars begin to appear
in R\,136 (see Figure \protect{\ref{fig_H301_R136}}). Hunter
et al.'s (1995) LF was normalized to coincide with $M_{V,0} = -2$ to
$-3$ magnitude bin of Hodge\,301 (arbitrarily chosen).  The brightest magnitude 
bins of Hodge\,301
are affected by small-number statistics and the lack of stars near the
main-sequence turnoff.  
\label{Fig_complete}}

\figcaption[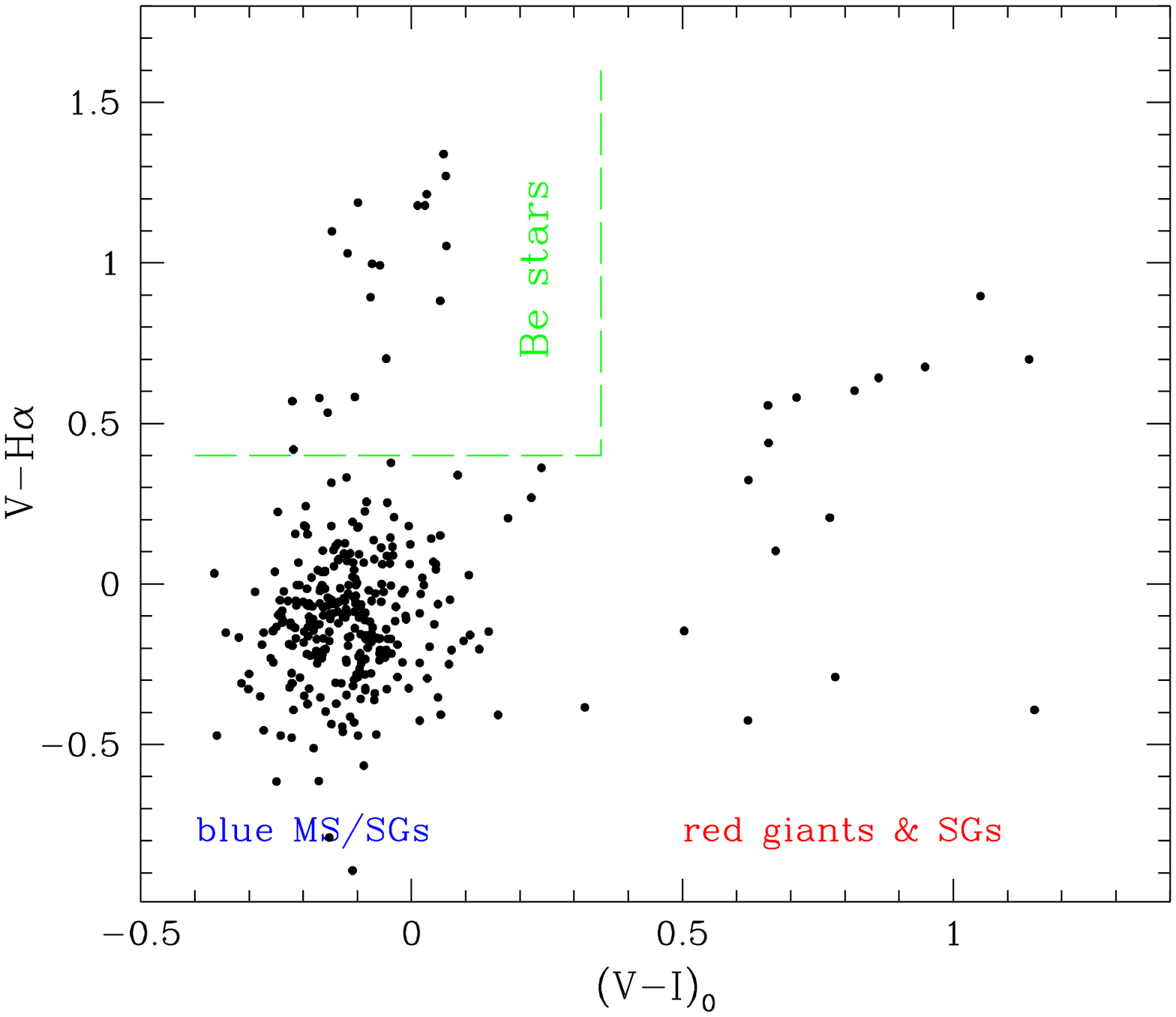]{A two-color diagram to identify stars with 
H$\alpha$ excess.  
The H$\alpha$ magnitudes are instrumental.  The bulk of main-sequence
stars (MS) and supergiants (SGs) shows a scattered distribution around
$(V-H\alpha) \sim 0$.  We consider blue stars with $(V-H\alpha) > 0\fm4$
to be Be star candidates.
\label{fig_TCD}}

\figcaption[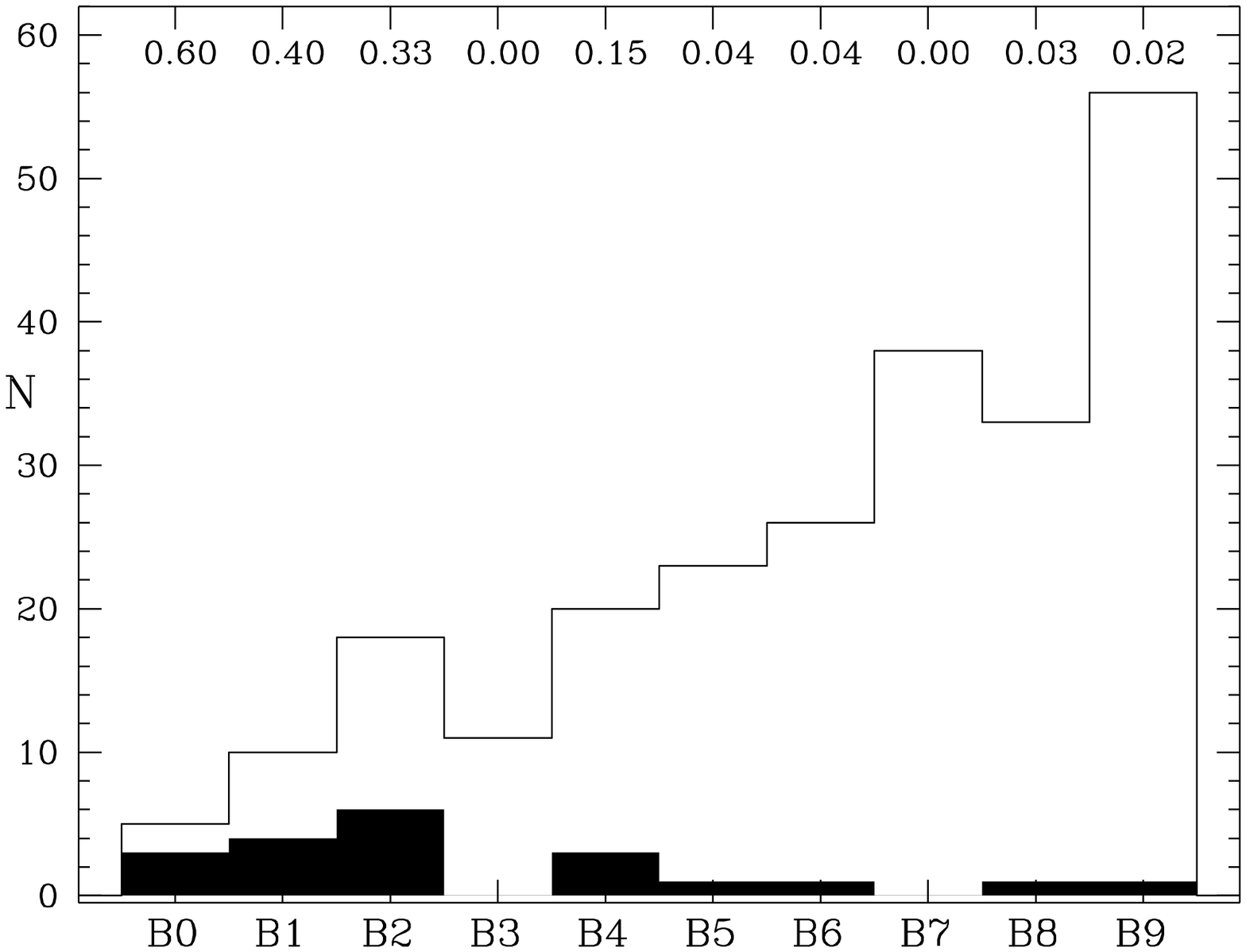]{Histogram of the number of Be star candidates 
(filled areas) and 
B main-sequence stars without emission (white areas) as a function of
photometrically estimated spectral type.  Numbers in the top row indicate
the number ratio Be/(Be+B).  The highest frequency of B emission star 
candidates is found among the early spectral types. 
\label{fig_Behist}}

\figcaption[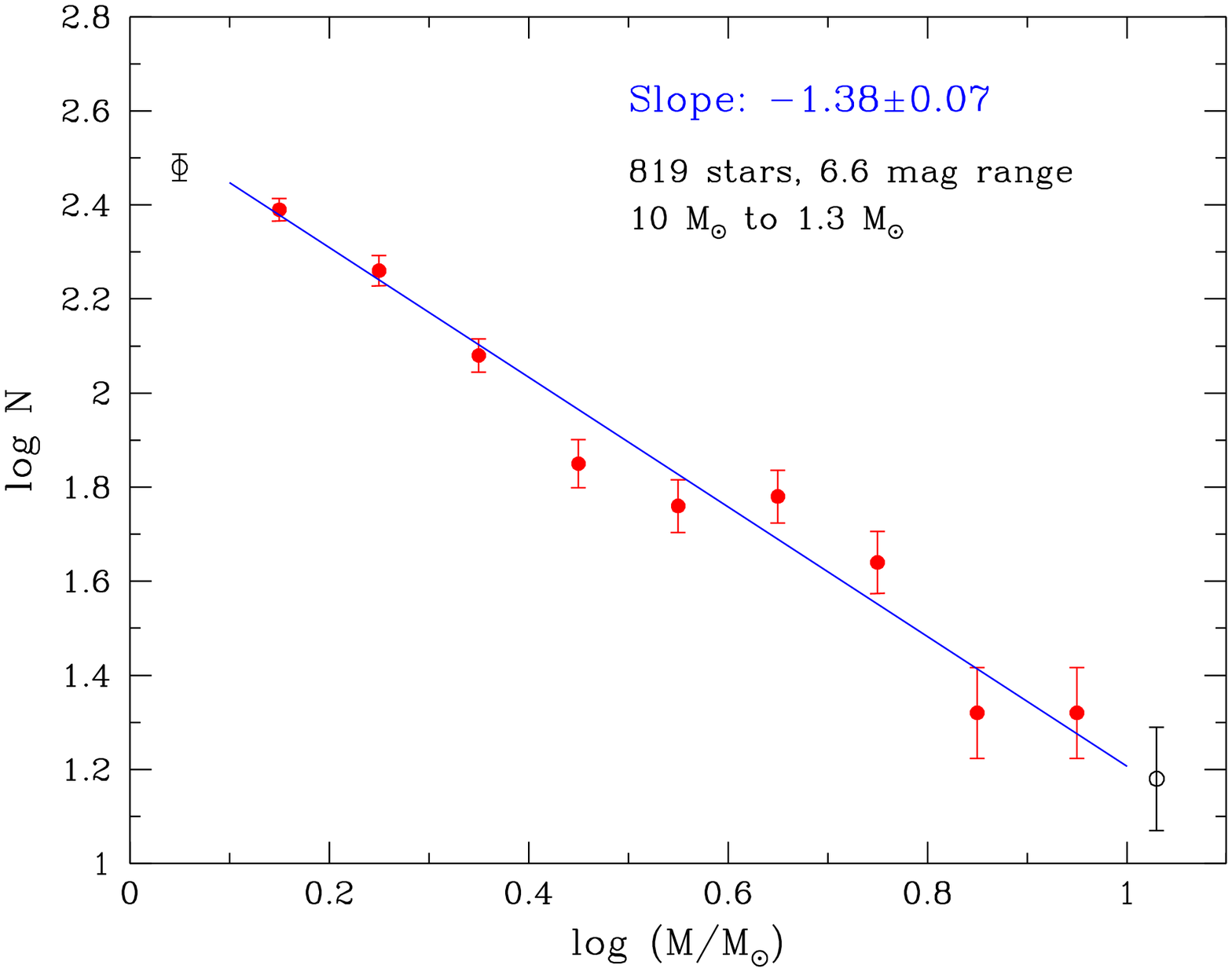]{The slope of Hodge\,301's mass function is 
$-1.38$ with a formal error of $\pm 0.07$, which is consistent with a Salpeter
slope.  The highest and the lowest mass bins (open circles) were excluded from
the calculation of the slope to minimize incompleteness and evolutionary
effects.
\label{fig_LF}}

\figcaption[grebel.fig9.eps]{Distribution of stellar generations of differing 
age in 30\,Dor.
The location of our {\em WFPC2} field is indicated.  Spectral classifications
were taken from Walborn \& Blades (1997), Parker (1993), and references 
therein. 
Upper left panel:  The oldest generation as traced by K, M and A supergiants
corresponding to ages of $\approx 20$ Myr.
The dot-dashed lines mark the approximate location of regions filled with
diffuse X-ray emission (after Wang 1999).  Upper right panel: Distribution of
B ($\approx > 10$ Myr) and O ($\approx > 3$ Myr) supergiants.  The dashed
circle encloses the R\,143 association, which contains mainly late O and 
early B supergiants (Walborn \& Blades 1997).  R\,143 itself is the only known
luminous blue variable in the 30\,Dor region (Parker et al.\ 1993) and believed
to be younger than 7 Myr (Walborn \& Blades 1997). 
Lower left panel:  Location of early O main-sequence stars and Wolf-Rayet stars
(ages $\approx <5$ Myr).  Note that many of the W-R stars are believed to be
very young main-sequence stars with ages of only 1 -- 2 Myr.  
R\,136 (dashed circle) contains a large
number of such W-R stars and O3 stars (Massey \& Hunter 1998), which are
not marked individually.
Lower right panel:  Some of the locations of embedded protostars and 
star-forming regions along
dense filaments (Walborn et al.\ 1999; Grebel et al., in prep.).
\label{fig_30Dor_age}}

\newpage

\begin{deluxetable}{lcccc}
\tablecaption{Log of the {\em HST} WFPC2 observations obtained on 1995 December 17.
\label{tbl_1}}
\tablehead{
\colhead{Filter}    & \colhead{Exposure time} & \colhead{Gain} & 
\colhead{CTE correction (PC1)} & \colhead{Image root name}\\
& [sec] & [e$^-$/ADU] & [\%] & }
\startdata
F656N & $2\times700$ &  7 & 2 & u3010101t, u3010102t \\
F555W & 4            & 15 & 4 & u3010103t \\
F555W & 40           &  7 & 4 & u3010104t \\
F555W & $2\times200$ &  7 & 4 & u3010105t, u3010106t \\
F336W & 10           & 15 & 4 & u3010109t \\
F336W & $2\times100$ &  7 & 4 & u301010at, u301010bt \\
F814W & 5            & 15 & 4 & u301010et \\
F814W & $2\times100$ &  7 & 4 & u301010ft, u301010gt \\
\enddata
\end{deluxetable}

\begin{deluxetable}{llllllllccc}
\tablecaption{Cross identifications, spectral types, and reddening 
of previously studied stars in Hodge 301.
\label{tbl_2}}
\tiny
\tablehead{
\colhead{MG73} & \colhead{MH81} & \colhead{Mk85} & 
\colhead{HSJG92} & \colhead{H93\tablenotemark{a}} & \colhead{WB97} & 
\colhead{SpT} & \colhead{UIT SpT} & $A_V$ MH81 & $E_{B-V,F}$ & $E_{B-V,FS}$ \\
\multicolumn{1}{c}{(1)} & \multicolumn{1}{c}{(2)} & \multicolumn{1}{c}{(3)} & 
\multicolumn{1}{c}{(4)} & \multicolumn{1}{c}{(5)} & \multicolumn{1}{c}{(6)} & 
\multicolumn{1}{c}{(7)} & \multicolumn{1}{c}{(8)} & \multicolumn{1}{c}{(9)} & 
\multicolumn{1}{c}{(10)} & \multicolumn{1}{c}{(11)} }
\startdata
MG\,91 & Dor\,IR\,4 & R\,132\tablenotemark{b} & HSJG\,4 & UIT\,269 & WB\,1 & B0.5\,Ia\tablenotemark{d} & B3-5\,I &      & 0.14 & 0.00 \\ 
       &            & B2  & HSJG\,6 &          & WB\,3 & A\,Ib\tablenotemark{d}    &         &      &      &      \\
MG\,93 & Dor\,IR\,35&     & HSJG\,9 &          & WB\,4 & M\,I\tablenotemark{c}     &         & 1.00 &      &      \\
MG\,94 & Dor\,IR\,7 &     & HSJG\,10&          & WB\,5 & M\,I\tablenotemark{c}     &        & 0.93 &      &      \\ 
       &            &     & HSJG\,7 & UIT\,264 & WB\,6 & B2\,III\tablenotemark{e}  & O7-B0\,V&      & 0.09 & 0.16 \\
MG\,95 & Dor\,IR\,8 &     & HSJG\,11&          & WB\,7 & M\,I\tablenotemark{c}     &         & 0.96 &     &       \\
       &            &     &         & UIT\,275 & WB\,8 & B5:\,p\tablenotemark{e}   &O9-B0\,VI&      & 0.15 & 0.24 \\
       &            &     &         & UIT\,250 & WB\,9 & B2\,V var\tablenotemark{e}&O5-6\,III&      & 0.11 & 0.38 \\
       &            & B1  & HSJG\,13&          & WB\,10& A0\,Ib\tablenotemark{d}   &         &      &      &      \\
       &            & B3  & HSJG\,16& UIT\,265 & WB\,11& A5\,Ib\tablenotemark{d}   &B1-1.5\,V&      & 0.09 & 0.09 \\
       &            &     &         & UIT\,241 &       &                           & O3-6\,V &      & 0.09 & 0.49 \\
       &            &     &         & UIT\,247 &       &                           & B3-5\,I &      & 0.16 & 0.02 \\
       &            &     &         & UIT\,256 &       &                           & B3-5\,I &      & 0.09 & 0.06 \\
       &            &     &         & UIT\,257 &       &                           &         &      & 0.16 & 0.01 \\
        &           &     &         & UIT\,258 &       &                           &O7-B0\,V &      & 0.09 & 0.15 \\
       &            &     &         & UIT\,259 &       &                           &O7-B0\,V &      & 0.09 & 0.14 \\
\tablenotetext{a}{Due to the large difference in resolution it is difficult
to unambiguously identify the {\em UIT} stars in our {\em HST} image.  The
cross identifications listed here represent our best guess.}
\tablenotetext{b}{Nomenclature from Feast et al.\ (1960).}
\tablenotetext{c}{Spectral type from McGregor \& Hyland (1981; MH81).}
\tablenotetext{d}{Spectral type from Melnick (1985; Mk85).)}
\tablenotetext{e}{Spectral type from Walborn \& Blades (1997; WB97).}
\enddata
\end{deluxetable}

\begin{deluxetable}{llccllllllllll}
\tablecaption{Coordinates and 
Photometry for Be Stars and Other Bright Stars in Hodge\,301.
\label{tbl_3}}
\tiny
\tablehead{
\colhead{H93\tablenotemark{a}} & \colhead{WB97} & \colhead{$\alpha$ 
(J2000)\tablenotemark{b}} &
\colhead{$\delta$ (J2000)\tablenotemark{b}} & Be star &
\colhead{$m_{162}$} & \colhead{$m_{189}$} & \colhead{$m_{221}$} &
\colhead{$m_{256}$} & \colhead{$U_0$} & \colhead{$B$} & \colhead{$V_0$} & 
\colhead{$I_0$} & \colhead{$K$}\\
\multicolumn{1}{c}{(1)} & \multicolumn{1}{c}{(2)} & \multicolumn{1}{c}{(3)} &
\multicolumn{1}{c}{(4)} & \multicolumn{1}{c}{(5)} & \multicolumn{1}{c}{(6)} &
\multicolumn{1}{c}{(7)} & \multicolumn{1}{c}{(8)} & \multicolumn{1}{c}{(9)} &
\multicolumn{1}{c}{(10)} & \multicolumn{1}{c}{(11)} & \multicolumn{1}{c}{(12)} 
& \multicolumn{1}{c}{(13)} & \multicolumn{1}{c}{(14)} }
\startdata
 UIT\,269 & WB\,1 & $\rm5^h38^m15^s$ & $-69^{\circ}03'47''$ &      & 12.26 & 12.46 & 12.24 & 12.41 &       & 13.05 &       & 12.26\tablenotemark{c} \\   
 UIT\,256 & WB\,2 & $\rm5^h38^m15.1^s$ & $-69^{\circ}04'00''$ & no   & 11.96 & 12.31 & 12.26 & 12.77 & 12.70 & 13.11 & 13.66 & 13.90 &  \\  
          & WB\,3 & $\rm5^h38^m15.4^s$ & $-69^{\circ}04'02''$ & no   &       &       &       &       & 12.82 &       & 12.12 & 11.94 & \\   
          & WB\,4 & $\rm5^h38^m16.7^s$ & $-69^{\circ}04'13''$ & no   &       &       &       &       & 16.17 &       & 13.40 & 11.16 & 8.39 \\   
          & WB\,5 & $\rm5^h38^m17.0^s$ & $-69^{\circ}04'00''$ & no   &       &       &       &       & 15.83 &       & 13.00 & 11.16 & 8.70 \\   
 UIT\,264 & WB\,6 & $\rm5^h38^m17.0^s$ & $-69^{\circ}03'49''$ & Be4  & 12.53 & 13.09 &       & 13.43 & 14.15 & 15.45 & 15.34 & 15.56 & \\  
          & WB\,7 & $\rm5^h38^m17.6^s$ & $-69^{\circ}04'11''$ & no   &       &       &       &       & 16.15 &       & 13.29 & 11.76 & 8.87 \\   
 UIT\,275 & WB\,8 & $\rm5^h38^m18.0^s$ & $-69^{\circ}03'38''$ &      & 13.61 & 13.69 &       & 14.39 &       & 15.11 &       &  & \\  
 UIT\,250 & WB\,9 & $\rm5^h38^m18.0^s$ & $-69^{\circ}04'08''$ & Be1  & 12.78 & 13.05 &       & 13.41 & 12.83 & 14.64 & 13.80 & 13.75 & \\  
          & WB\,10& $\rm5^h38^m18.3^s$ & $-69^{\circ}04'01''$ & no   &       &       &       &       & 12.71 &       & 12.49 & 12.45 & \\   
 UIT\,265 & WB\,11& $\rm5^h38^m22.0^s$ & $-69^{\circ}03'47''$ &      & 13.71 &       &       & 15.01 &       & 16.92 &       & & \\   
 UIT\,257 &       & $\rm5^h38^m18.8^s$ & $-69^{\circ}03'59''$ & Be3  & 12.44 & 12.87 & 11.91 & 13.00 & 14.22 & 13.36 & 15.27 & 15.24 & \\   
 UIT\,259 &       & $\rm5^h38^m16.0^s$ & $-69^{\circ}03'53''$ & Be2  & 12.42 & 12.68 &       & 13.58 & 13.90 & 15.50 & 15.15 & 15.21 & \\  
 UIT\,247 &       & $\rm5^h38^m19.7^s$ & $-69^{\circ}04'09''$ & no   &       & 14.31 &       & 14.73 & 14.34 & 16.25 & 15.50 & 15.66 & \\  
 UIT\,241 &       & $\rm5^h38^m16.5^s$ & $-69^{\circ}04'13''$ & no   & 13.66 &       &       & 14.62 & 14.90 & 15.53 & 16.06 &  16.32 & \\   
 UIT\,258 &       & $\rm5^h38^m17.0^s$ & $-69^{\circ}04'01''$ & no   & 12.61 & 13.04 &       & 13.33 & 14.36 & 15.23 & 15.53 & 15.77 & \\   
          &       & $\rm5^h38^m15.6^s$ & $-69^{\circ}04'01''$ & Be5  &       &       &       &       & 14.18 &       & 15.35 & 15.29 & \\
          &       & $\rm5^h38^m18.7^s$ & $-69^{\circ}03'58''$ & Be6  &       &       &       &       & 14.31 &       & 15.48 & 15.65 & \\
          &       & $\rm5^h38^m16.8^s$ & $-69^{\circ}03'56''$ & Be7  &       &       &       &       & 14.88 &       & 15.70 & 15.78 & \\
          &       & $\rm5^h38^m16.4^s$ & $-69^{\circ}03'50''$ & Be8  &       &       &       &       & 14.86 &       & 16.01 & 15.98 & \\
          &       & $\rm5^h38^m17.3^s$ & $-69^{\circ}03'55''$ & Be9  &       &       &       &       &       &       & 16.25 & 16.19 & \\
          &       & $\rm5^h38^m17.0^s$ & $-69^{\circ}03'50''$ & Be10 &       &       &       &       &       &       & 16.32 & 16.31 & \\
          &       & $\rm5^h38^m17.3^s$ & $-69^{\circ}04'01''$ & Be11 &       &       &       &       &       &       & 16.45 & 16.57 & \\
          &       & $\rm5^h38^m16.8^s$ & $-69^{\circ}03'54''$ & Be12 &       &       &       &       & 15.59 &       & 16.68 & 16.83 & \\
          &       & $\rm5^h38^m15.7^s$ & $-69^{\circ}03'53''$ & Be13 &       &       &       &       & 15.63 &       & 16.73 & 16.67 & \\
          &       & $\rm5^h38^m18.7^s$ & $-69^{\circ}04'03''$ & Be14 &       &       &       &       & 16.74 &       & 17.20 & 17.24 & \\
          &       & $\rm5^h38^m15.7^s$ & $-69^{\circ}03'47''$ & Be15 &       &       &       &       & 16.37 &       & 17.36 & 17.46 & \\
          &       & $\rm5^h38^m17.6^s$ & $-69^{\circ}03'56''$ & Be16 &       &       &       &       & 16.61 &       & 17.53 & 17.64 & \\
          &       & $\rm5^h38^m15.5^s$ & $-69^{\circ}03'56''$ & Be17 &       &       &       &       & 16.86 &       & 17.69 & 17.76 & \\
          &       & $\rm5^h38^m15.9^s$ & $-69^{\circ}03'59''$ & Be18 &       &       &       &       & 17.23 &       & 18.10 & 18.26 & \\
          &       & $\rm5^h38^m16.3^s$ & $-69^{\circ}03'57''$ & Be19 &       &       &       &       &       &       & 18.66 & 18.88 & \\
\tablenotetext{a}{Due to the large difference in resolution it is difficult
to unambiguously identify the {\em UIT} stars in our {\em HST} image.  The
cross identifications listed here represent our best guess.}
\tablenotetext{b}{The equatorial coordinates were determined with the task
{\sc metric} in STSDAS on the PC image.  The coordinates for the stars 
WB1, WB8, and WB11 were determined on images from the Digitized Sky Survey
using the IRAF/PROS task {\sc tvlabel}.}
\tablenotetext{c}{Undereddened $I$ magnitude from Mendoza \& G\'omez (1973).}
\enddata
\end{deluxetable}

\begin{deluxetable}{rcrc}
\tablecaption{Mass range, magnitude bins\tablenotemark{1} 
and main-sequence star counts for Hodge\,301.
\label{tbl_4}}
\tablehead{
\colhead{Mass range ($M_{\odot}$)} & \colhead{log ($M/M_{\odot}$)} & 
\colhead{$M_V$} & \colhead{$N$} }
\startdata
10.0 -- 7.9 & 1.0 &  -2.90  & $ 21\pm4.6$  \\
 7.9 -- 6.3 & 0.9 &  -1.83  & $ 21\pm4.6$  \\
 6.3 -- 5.0 & 0.8 &  -1.11  & $ 44\pm6.6$  \\
 5.0 -- 4.0 & 0.7 &  -0.51  & $ 60\pm7.7$  \\
 4.0 -- 3.2 & 0.6 &   0.07  & $ 57\pm7.5$  \\
 3.2 -- 2.5 & 0.5 &   0.65  & $ 71\pm8.4$  \\
 2.5 -- 2.0 & 0.4 &   1.24  & $119\pm10.9$  \\
 2.0 -- 1.6 & 0.3 &   1.90  & $183\pm13.5$  \\
 1.6 -- 1.3 & 0.2 &   2.64  & $243\pm15.6$  \\
 1.3 -- 1.0 & 0.1 &   3.74  & $305\pm17.5$  \\
\enddata
\tablenotetext{1}{Logarithmic mass and magnitude entries denote the upper 
boundaries of each mass and magnitude bin. 
}
\end{deluxetable}


\begin{thebibliography}{}

\bibitem{} Bertelli, G., Bressan, A., Chiosi, C., Fagotto, F., \& Nasi, E.
1994, A\&AS, 106, 275 

\bibitem{} Bica, E.L.D., Schmitt, H.R., Dutra, C.M., Oliveira, \& H.L. 1999,
AJ, 117, 238

\bibitem{} Bohlin, R.C., Savage, B.D., \& Drake, J.F. 1978, ApJ, 244, 132

\bibitem{} Brandl, B., Sams, B.J., Bertoldi, F., Eckart, A., Genzel, R., 
Drapatz, S., Hofmann, R., L\"owe, M., \& Quirrenbach, A. 1996, ApJ, 466, 254

\bibitem{} Chu, Y.-H., \& Kennicutt, R.C., Jr. 1994, ApJ, 425, 720

\bibitem{} Collins, G.W., \& Smith, R.C. 1985, MNRAS, 213, 519 

\bibitem{} Collins, G.W., Truax, R.J., \& Cranmer, S.R. 1991, ApJS, 77, 541

\bibitem{} Cox, P., \& Deharveng, L. 1983, A\&A, 117, 265

\bibitem{} Crotts, A.P.S., Kunkel, W.E., \& Heathcote, S.R. 1995, ApJ, 438, 724

\bibitem{} Elmegreen, B.G. 1997, ApJ, 486, 944

\bibitem{} Fabregat, J., \& Torrejon, J.M. 1999, A\&A, submitted 
(astro-ph/9906240)

\bibitem{} Feast, M.W., Thackeray, A.D., \& Wesselink, A.J. 1960, MNRAS, 121,
25

\bibitem{} Fitzpatrick, E.L. 1985, ApJ, 299, 219 (F)

\bibitem{} Fitzpatrick, E.L. 1986, AJ, 92, 1068

\bibitem{} Fitzpatrick, E.L., \& Savage, B.D. 1984, ApJ, 279, 578 (FS)

\bibitem{} Grebel, E.K. 1997, A\&A, 317, 448

\bibitem{} Grebel, E.K., \& Chu, Y.-H. 1996, AGAb, 12, 215

\bibitem{} Grebel, E.K., Richtler, T., de Boer, K.S. 1992, A\&A, 254, L5

\bibitem{} Grebel, E.K., \& Roberts, W.J. 1995, A\&AS, 109, 293

\bibitem{} Grebel, E.K., Roberts, W.J., \& Brandner, W. 1996, 311, 470 

\bibitem{} Grebel, E.K., Roberts, W.J., Will, J.M., \& de Boer, K.S. 1993, 
SSRv, 66, 65

\bibitem{} Heap, S.R., Altner, B., Ebbets, D., Hubeny, I., Hutchings, J.B.,
Kudritzki, R.P., Voels, S.A., Haser, S., Pauldrach, A., Puls, J., \&
Butler, K. 1991, ApJ, 377, L29

\bibitem{} Hill, J.K., Bohlin, R.C., Cheng, K.-P., Fanelli, M.N., Hintzen, P.,
O'Connell, R.W., Roberts, M.S., Smith, A.M., Smith, E.P., \& Stecher, T.P.
1993, ApJ, 413, 604

\bibitem{} Hodge, P. 1988, PASP, 100, 1051

\bibitem{} Holtzman, J., Burrows, C.J., Casertano, S., Hester, J.J., 
Trauger, J.T., Watson, A.M., \& Worthey, G. 1995, PASP, 107, 1065

\bibitem{} Hunter, D.A., Shaya, E.J., Holtzman, J.A., Light, R.M., 
O'Neil, E.J., \& Lynds, R. 1995, ApJ, 448, 179

\bibitem{} Hunter, D.A., O'Neil, E.J., Lynds, R., Shaya, E.J., Groth, E.J., 
\& Holtzman, J.A. 1996, ApJ, 459, L27

\bibitem{} Hunter, D.A., Light, R.M., Holtzman, J.A., Lynds, R., O'Neil, E.J., 
\& Grillmair, C.J. 1997b, ApJ, 478, 124

\bibitem{} Hunter, D.A., Vacca, W.D., Massey, P., Lynds, R., \& O'Neil, E.J.
           1997a, AJ, 113, 1691

\bibitem{} Hyland, A.R., Thomas, J.A., \& Robinson, G. 1978, AJ, 83, 20

\bibitem{} Hyland, A.R., Straw, S., Jones, T.J., \& Gatley, I. 1992, MNRAS, 257,
391

\bibitem{} Miller, G.E., \& Scalo, J.M. 1979, ApJS, 41, 513

\bibitem{} Keller, S.C., Wood, P.R., \& Bessell, M.S. 1999, A\&AS, 134, 489

\bibitem{} Kennicutt, R.C. 1984, ApJ, 287, 116

\bibitem{} Keyes, T. 1997, HST Data Handbook Vers.\ 3, Space Telescope
Science Institute, Baltimore

\bibitem{} Lortet, M.-C., \& Testor, G. 1991, A\&AS, 89, 185

\bibitem{} Maeder, A., Grebel, E.K., \& Mermilliod, J.-C. 1999, A\&A, 346, 459 

\bibitem{} Malumuth, E.M., \& Heap, S.R. 1994, AJ, 107, 1054

\bibitem{} Massey, P., Johnson, K.E., \& DeGoia-Eastwood, K. 1995b, ApJ, 454, 151

\bibitem{} Massey, P., Lang, C.C., DeGoia-Eastwood, K., \& Garmany, C.D. 1995a,
ApJ,  438, 188

\bibitem{} Massey, P., \& Hunter, D.A. 1998, ApJ, 493, 180

\bibitem{} McGregor, P.J., \& Hyland, A.R. 1981, ApJ, 250, 116

\bibitem{} Meaburn, J. 1988, MNRAS, 235, 375

\bibitem{} Melnick, J. 1985, A\&A, 153, 235

\bibitem{} Mendoza, E.E., \& G\'omez, T. 1973, PASP, 85, 439

\bibitem{} Norci, L., \& \"Ogelman, H. 1995, A\&A, 302, 879

\bibitem{} Panagia, N. 1999, in `New Views of the Magellanic Clouds'', IAU
Symp.\ 190, eds.\ Y.-H.\ Chu et al., ASP Conf. Ser., in press 

\bibitem{} Parker, J.W. 1993, AJ, 106, 560

\bibitem{} Parker, J.W., \& Garmany, C.D. 1993, AJ, 106, 1471

\bibitem{} Parker, J.W., Clayton, G.C., Winge, C., \& Conti, P.S. 1993,
ApJ, 409, 770

\bibitem{} Parker, J.W., Hill, J.K., Cornett, R.H., Hollis, J., Zamkoff, E.,
Bohlin, R.C., O'Connell, R.W., Neff, S.G., Roberts, M.S., Smith, A.M., \&
Stecher, T.P. 1998, AJ, 116, 180

\bibitem{} Pinfield, D.J., Jameson, R.F., \& Hodgkin, S.T. 1998, MNRAS, 299, 955

\bibitem{} Pols, O.R., \& Marinus, M. 1994, A\&A, 288, 475

\bibitem{} Raboud, D., \& Mermilliod, J.-C., 1998, A\&A, 333, 897

\bibitem{} {\em ROSAT} Mission Description 1991, NASA Publication NRA 91-OSSA-25, 
 Appendix F

\bibitem{} Rubio, M., Barb\'a, R.H., Walborn, N.R., Probst, R.G., Garc\'{\i}a,
J., \& Roth, M.R. 1998, AJ, 116, 1708

\bibitem{} Rubio, M., Roth, M., \& Garc\'{\i}a, J. 1992, A\&A 261, L29

\bibitem{} Sagar, R., \& Griffiths, W.K. 1998, MNRAS, 299, 777

\bibitem{} Salpeter, E.E. 1955, ApJ, 121, 161

\bibitem{} Scalo, J.M. 1986, Fundam.\ Cosmic Phys., 11, 1

\bibitem{} Schaerer, D., Meynet, G., Maeder, A., \& Schaller, G.
1993, A\&AS, 98, 523 

\bibitem{} Selman, F., Melnick, J., Bosch, G., \& Terlevich, R. 1999,
A\&A, 347, 532

\bibitem{} Sirianni, M., Leitherer, C., Nota, A., Clampin, M., \& de Marchi,
G. 1999, in
IAU Symp.\ 190, ``New views of the Magellanic Clouds'', eds.\ Y.-H.\ Chu
et al., ASP Conf.\ Ser., in press

\bibitem{} Stecher, T. P., et al. 1992, ApJ, 395, L1

\bibitem{} Stetson, P.B. 1992, in `Astronomical Data Analysis Software and
Systems I', ASP Conf.\ Ser.\ 25, eds.\ D.M. Worrall, C. Biemesderfer, \&
J.\ Barnes, p. 297

\bibitem{} Trauger, J.T., et al. 1994, ApJ, 435, L3

\bibitem{} Thornley, M.D., Lutz, D., Kunze, D., \& Spoon, H. 1999, in
IAU Symp.\ 190, ``New views of the Magellanic Clouds'', eds.\ Y.-H.\ Chu
et al., ASP Conf.\ Ser., in press

\bibitem{} van den Bergh, S. 1999, in
IAU Symp.\ 190, ``New views of the Magellanic Clouds'', eds.\ Y.-H.\ Chu
et al., ASP Conf.\ Ser., in press

\bibitem{} Walborn, N.R. 1991, in IAU Symp.\ 148 `The Magellanic Clouds', eds.\
R.\ Haynes \& D.\ Milne, Kluwer, Dordrecht, p.\ 145

\bibitem{} Walborn, N.R., \& Blades, J.C. 1987, ApJ, 323, L65

\bibitem{} Walborn, N.R., \& Blades, J.C. 1997, ApJS, 112, 457

\bibitem{} Walborn, N.R., Barb\'a, R.H., Brandner, W., Rubio, M., Grebel,
           E.K., \& Probst, R.G. 1999, AJ, 117, 225

\bibitem{} Wang, Q.D. 1995, ApJ, 453, 783 

\bibitem{} Wang, Q.D. 1999, ApJ, 510, L139

\bibitem{} Wang, Q.D., Helfand, D.J. 1991, ApJ, 370, 541

\bibitem{} Westerlund, B.E. 1961, Uppsala Astron.\ Obs.\ Annals, 5, \# 1

\bibitem{} Westerlund, B.E. 1997, ``The Magellanic Clouds'', Cambridge 
University Press, p.\ 20

\bibitem{} Yang, H., Chu, Y.-H., Skillman, E.D., \& Terlevich, R. 1996,
112, 146

\bibitem{} Zorec, J., \& Briot, D. 1991, A\&A, 245, 150

\end{thebibliography}
\end{document}